\documentstyle[epsfig]{article}
\newcommand{\w}{\omega}
\newcommand{\prps}{\left\langle\frac{\partial R}{\partial
\theta_s}\right\rangle}
\newcommand{\p}{\partial}
\newcommand{\ts}{\theta_s}
\newcommand{\la}{\langle}
\newcommand{\ra}{\rangle}
\renewcommand{\to}{\theta_0}
\newcommand{\iii}{\int^{\infty}_{-\infty}}
\newcommand{\zx}{\zeta (x_1)}
\newcommand{\bqe}{\begin{eqnarray}}
\newcommand{\eqe}{\end{eqnarray}}
\begin{document}

\begin{center}
{\bf  Design of one-dimensional Lambertian diffusers
of light}\vspace*{.2in}\\
A. A. Maradudin$^*$, I. Simonsen$^{*,\dag}$, T. A. Leskova$^{\ddag}$,
and E. R. M\'endez$^{\natural}$\\
$^*${\it Department of Physics and Astronomy and Institute for
Surface and Interface Science, University of California, Irvine, CA
92697, U.S.A.}\\
$^{\dag}${\it Department of Physics, The Norwegian University of
Science and Technology, N-7491, Trondheim, Norway}\\
$^{\ddag}${\it Institute of Spectroscopy, Russian Academy of
Sciences, Troitsk, Russia}\\
$^{\natural}${\it Divisi\'on de F\'{\i}sica Aplicada, CICESE,
Ensenada, Baja California 22800, M\'exico}\\
\bigskip
\bigskip
{\bf Abstract}
\end{center}
We describe a method for designing a one-dimensional random surface
that acts as a Lambertian diffuser.  The method is tested by means of
rigorous  computer simulations and is shown to yield the desired
scattering pattern.
\newpage

Optical devices that give rise to a scattered intensity that is
proportional to the cosine of the scattering angle are frequently
used in the optical industry, e.g. for calibrating scatterometers
[1].  Such diffusers have the property that their radiance or
luminance is the same in {\it all} scattering directions.  Due to
this angular dependence such devices are often referred to as {\it
Lambertian diffusers}.    In the visible region of the optical
spectrum volume disordered media, e.g. compacted powdered  barium 
sulphate, and freshly smoked magnesium oxide [2]
are used as Lambertian diffusers.  However, this type of diffuser
is inapplicable in the infrared region due to its strong absorption
and the presence of a specular component in the scattered light, in
this frequency range.

The design of a random surface that acts as a Lambertian diffuser,
especially in the infrared region of the optical spectrum, is
therefore a desirable goal, and one that has been regarded as
difficult to achieve [3].  In this paper we present a solution to this
problem that is based on an approach used in several recent papers to
design one-dimensional random surfaces with specified scattering
properties [4-6], and to fabricate them in the laboratory [5,7].  The
design of a two-dimensional random surface that acts as a Lambertian
diffuser will be described elsewhere [8].

To motivate the calculations that follow we begin by considering the
scattering of s-polarized light of frequency $\w$ from a
one-dimensional, randomly rough, perfectly conducting surface defined
by $x_3 = \zx$.  The region $x_3 > \zx$ is vacuum, the region $x_3 <
\zx$ is the perfect conductor (Fig. 1).
\begin{figure}[htb]
\begin{center}
\epsfig{file=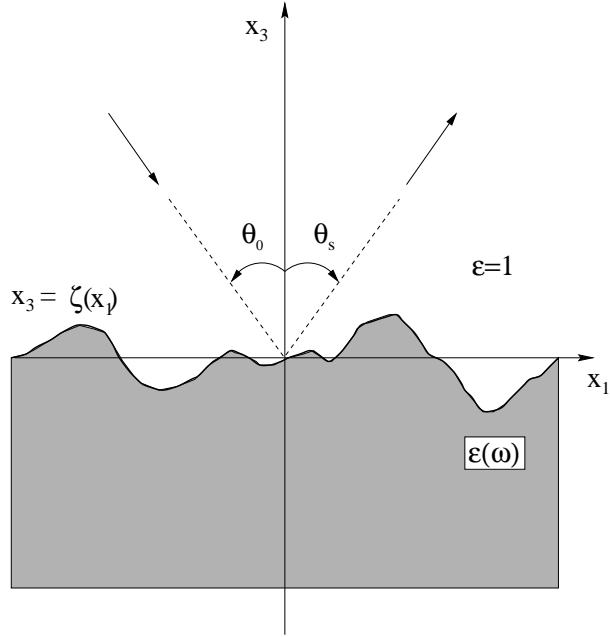,width=8cm}
\caption{ The scattering geometry assumed in this paper.}
\end{center}
\end{figure}
The plane of incidence is
the $x_1x_3$-plane.  The surface profile function  $\zx$ is
assumed to be a single-valued function of $x_1$ that is
differentiable, and to constitute a random process.

The mean differential reflection coefficient $\la \p R/\p\ts\ra$,
where the angle brackets denote an average over the ensemble of
realizations of the surface profile function, is defined such that
$\la \p R/\p\ts\ra d\ts$ is the fraction of the total time-averaged
flux incident on the surface that is scattered into the angular
interval $(\ts ,\ts + d\ts )$ in the limit as $d\ts \rightarrow 0$.
In the geometrical optics limit of the Kirchhoff approximation it is
given by [5]
\bqe
\prps &=& \frac{1}{L_1} \frac{\w}{2\pi c}
\frac{1}{\cos\to} \left[ \frac{1+\cos (\to + \ts )}{\cos\to +
\cos\ts}\right]^2 \iii dx_1\iii du \, \exp [i(q-k)u]\nonumber\\
&& \hspace*{1.2in} \times \la \exp [iau\zeta '(x_1)]\ra .
\eqe
In this expression $L_1$ is the length of the $x_1$-axis covered by 
the random surface, $\to$ and $\ts$ are the angles of incidence and
scattering, respectively, $a = (\w /c)(\cos\to + \cos\ts )$, and $q =
(\w /c)\sin\ts , k = (\w /c )\sin\to$.  In what follows, we will
restrict ourselves to the case of normal incidence $(\to =
0^{\circ})$, in which case Eq. (1) simplifies to
\bqe
\prps = \frac{1}{L_1} \frac{\w}{2\pi c} \iii dx_1\iii du \, \exp iqu
\la \exp [iau\zeta '(x_1)]\ra ,
\eqe
where $a$ is now given by $a = (\w /c) (1+\cos\ts )$.

We wish to find a surface profile function $\zx$ for which the mean
differential reflection coefficient has the form
\bqe
\prps = \frac{1}{2}\cos\ts .
\eqe
To this end we write $\zx$ in the form [5]
\bqe
\zx = \sum^{\infty}_{\ell = -\infty} c_{\ell}\, s(x_1-\ell 2b) .
\eqe
Here the $\{ c_{\ell}\}$ are independent, positive, random deviates,
$b$ is a characteristic length, and the function $s(x_1)$ is defined
by [5]
\bqe
s(x_1) = \left\{ \begin{array}{ccc}
0 & & x_1 \leq - (m+1)b\\
-(m+1)bh-hx_1 & \,\,\, & -(m+1)b\leq x_1 \leq -mb\\
-bh & & -mb \leq x_1 \leq mb\\
-(m+1)bh+hx_1 & & mb\leq x_1 \leq (m+1)b\\
0 & & (m+1)b \leq x_1
\end{array}\right. ,
\eqe
where $m$ is a positive integer.  Such trapezoidal grooves can be
generated experimentally [5,7].

Since the $\{ c_{\ell}\}$ are positive random deviates, their
probability density function (pdf) $f(\gamma ) = \la \delta (\gamma
-c_{\ell})\ra$ is nonzero only for positive values of $\gamma$.

It has been shown [5] that when the surface profile function is given
by Eqs. (4) and (5), the expression (2) for the mean differential
reflection  coefficient becomes
\bqe
\prps = \frac{1}{4h} \left(1+\tan^2\frac{\ts}{2}\right)\left[ f\left(
- \frac{1}{h}\tan\frac{\ts}{2}\right) + f\left(
\frac{1}{h}\tan\frac{\ts}{2}\right)\right] .
\eqe
Thus, we find that in the geometrical optics limit of the Kirchhoff
approximation the mean differential reflection  coefficient  is
determined by the pdf of the coefficients $\{ c_{\ell}\}$ entering
the expansion (4), and is independent  of the wavelength of the
incident light.  If we make the change of variable $\tan (\ts /2) =
\gamma h, 0 \leq \gamma h \leq 1$, so that $\frac{1}{2} \cos\ts =
\frac{1}{2} (1-\gamma^2 h^2)/(1+\gamma^2h^2)$, on combining Eqs. (3)
and (6) we find that the equation determining $f(\gamma )$ is
\bqe
f(-\gamma ) + f(\gamma ) = 2h \frac{1-\gamma^2h^2}{(1+\gamma^2h^2 )^2} .
\eqe
It follows that
\bqe
f(\gamma ) = 2h \frac{1-\gamma^2h^2}{(1+\gamma^2h^2)^2} \theta
(\frac{1}{h}-\gamma)\theta (\gamma ) .
\eqe

The preceding results were obtained in the geometrical optics limit
of the Kirchhoff approximation for a perfectly conducting surface.
However, our earlier  experience  in designing surfaces with
specified scattering properties [4-6] shows that when a surface
designed on the basis of these assumptions is ruled on a lossy metal,
the results of rigorous scattering calculations show that the
resulting scattering pattern retains the form prescribed in the
approximate, single-scattering calculations.  We now demonstrate that
such a result is obtained in the context of the present problem.

From the form of $f(\gamma )$ given in Eq. (8) a long sequence of $\{
c_{\ell}\}$ was generated by applying the rejection method [9], and
the resulting surface profile function $\zx$ was generated by the use
of Eqs. (4) and (5).  We found from numerical experiments that in
order to have a surface that acts as a Lambertian diffuser in
reflection the parameter $b$ had to be large.  Physically this means
that the grooves $\zx$ have to be wide.

In Fig.~2 we present the results of rigorous numerical Monte Carlo
\begin{figure}[htb]
\begin{center}
\epsfig{file=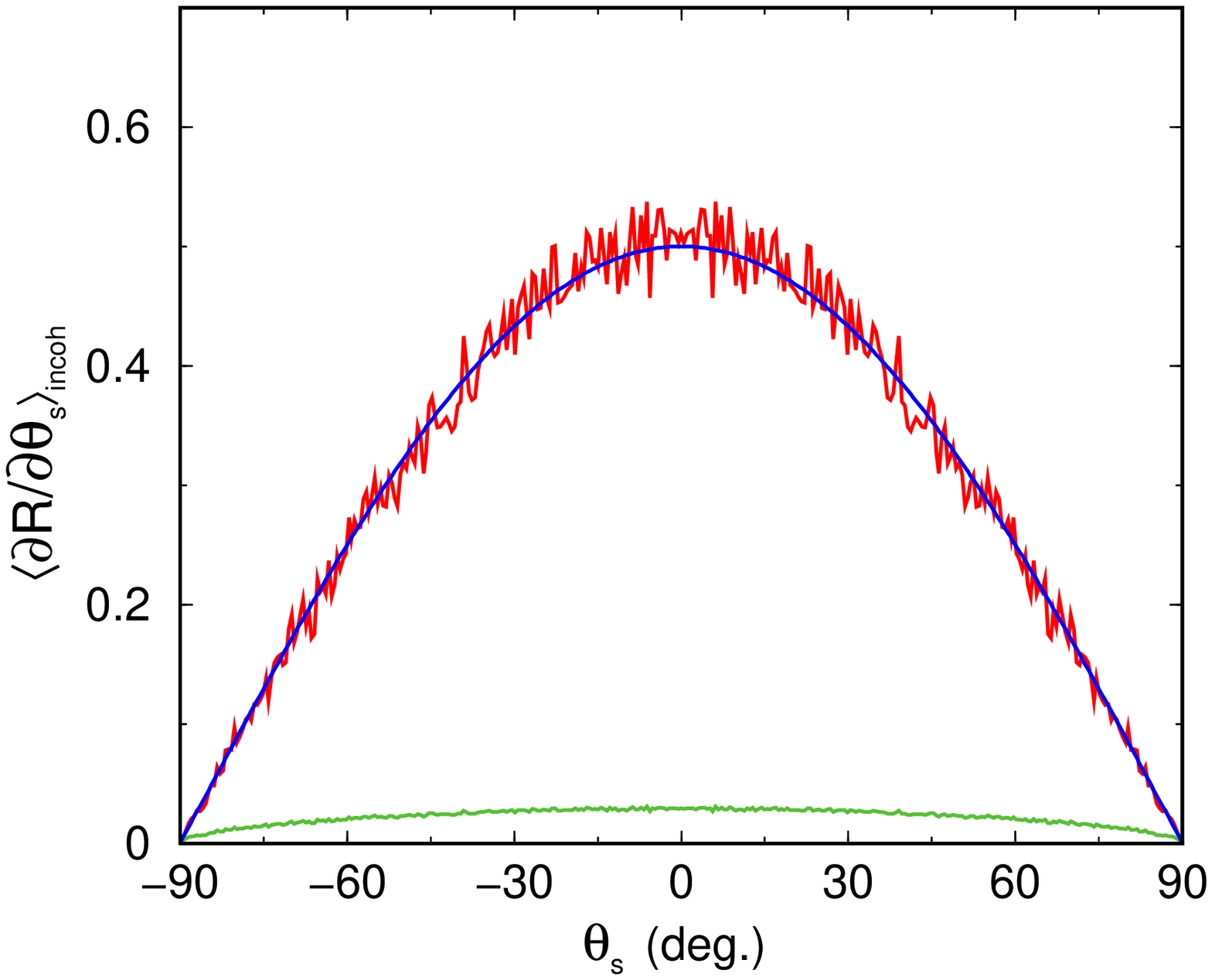,width=8cm}
\end{center}
\caption{ The noisy curve is $\la \p R/\p\ts\ra$ calculated
by a numerical simulation approach for a random silver surface
defined  by Eqs. (4) and (5) with $b = 80\lambda$, $h = 0.2$,  $m
= 1$, and the pdf (8), when s-polarized light of wavelength $\lambda
= 612.7\mbox{nm}$ $(\epsilon (\w ) = - 17.2 + i0.5)$ is incident normally
on it.  The upper solid curve is  $\la \p R/\p\ts\ra$ given by
Eq.~(3). The lower solid curve is the error in the calulated mean
differential reflection coefficient as measured by its standard deviation.}
\end{figure}
simulations [10] for the angular dependence of the mean differential
reflection coefficient $\la \p R/\p\ts \ra$ for s-polarized incident
light of wavelength $\lambda = 612.7\mbox{nm}$ scattered from a randomly
rough silver surface of the type described above (noisy curve).  The
value of the dielectric constant of silver at this wavelength is
$\epsilon (\w ) = - 17.2 +i0.5$.  The surface was characterized
by the parameters $b = 80\lambda$ $= 49\mu$m, $h = 0.2$, and $m = 1$,
and its length used in the simulation was $L_1 = 164\lambda =
100\mu$m.  Furthermore, the plot in Fig. 2 was obtained by averaging
the results for $N_{\zeta} = 35,000$ realizations of the surface
profile function $\zx$.  Such a large number of surface realizations
was needed  in order to reduce the noise level sufficiently.  The
reason for the slow convergence of the mean DRC with increasing $N_{\zeta}$
we believe is due to the large value of $b$ used in the simulations.
Without compromising the spatial discretization used in the numerical
calculation $(\Delta x_1 = 0.164\lambda )$ needed  in order to
resolve the oscillations of the incident field, only a few grooves
$s(x_1)$ could be included for each realization in the sum (4)
defining  the surface.

The lower smooth curve represents an estimate of the error in the
calculated $\la \p R/\p\ts \ra$ due to the use of an finite number of
surface realizations for its calculation. This error is obtained as
the standard deviation of the mean differential reflection
coefficient (see Ref.~[10] for details).

The upper smooth solid curve in Fig.~2 represents the geometrical
optics limit of the Kirchhoff approximation, Eq.~(3).  As can be
readily observed from this figure, the agreement between the
geometrical optics limit of the Kirchhoff approximation for a random
perfectly conducting surface and the result of rigorous numerical
simulations for a real random silver surface is excellent within the
noise level.  This is indeed the case for all scattering angles $\ts$,
which we find somewhat surprising, since one might have expected the
geometrical optics approximation to break down for the largest
scattering angles.  That this is not observed in our simulation
results is probably an indication that multiple scattering processes
are of minor importance in the scattering taking place at the random
surface even for the largest scattering angles.

Simulations (results not shown) were also performed where the
wavelength of the incident light was changed by plus and minus $10\%$
from its original value of $\lambda = 612.7\mbox{nm}$.  Such changes
did not affect the Lambertian nature of the scattered light in any
significant way. This weak wavelength sensitivity is consistent with
our earlier experience in designing surfaces with specified scattering
properties [4-6]. Surfaces generated on the basis of different $b$
parameters have also been considered. We found that the scattered
intensity showed little sensitivity to this parameter as long as it is
large.

\bigskip
\noindent
{\bf Acknowledgments}

The work of A.A.M. and T.A.L. was supported by Army Research Office
Grant DAAD 19-99-1-0321.  I.S. would like to thank the Research
Council of Norway (Contract No. 32690/213) and Norsk Hydro ASA for
financial support.  The work of E.R.M. was supported by CONACYT Grant
3804P-A.  This work has also received support from the Research
Council of  Norway (Program for Supercomputing) through a grant of
computing time.
\newpage
\noindent{\bf References}
\begin{description}
\item{[1]}
Stover J C 1995 {\it Optical Scattering} (Bellingham, WA, USA : SPIE Press)
\item{[2]}
Grum F and Luckey G W 1968 Optical sphere paint and a working 
standard of reflectance {\it Appl. Opt.} {\bf 7} 2295-2300 (1968).
\item{[3]}
Baltes H P 1980 Progress in inverse optical problems, in {\it Inverse 
Scattering Problems in Optics} ed. H.
P. Baltes (New York : Springer-Verlag) pp 1-13
\item{[4]}
Leskova T A, Maradudin A A, Novikov I V, Shchegrov A V and
M\'endez E R 1998
Design of one-dimensional band-limited uniform diffusers of light 
{\it Appl. Phys. Lett.} {\bf 73} 1943-1945
\item{[5]}
M\'endez E R,  Martinez-Niconoff G, Maradudin A A and
Leskova T A  1998 Design and synthesis of random uniform diffusers
{\it SPIE} {\bf 3426} 2-13
\item{[6]}
Maradudin A A, Simonsen I, Leskova T A and M\'endez E R 1999
Random surfaces that suppress single scattering
{\it Opt. Lett.} {\bf 24} 1257-1259
\item{[7]}
Chaikina E I, Garc\'{\i}a-Guerrero E E,  Gu Z-H,
Leskova T A, Maradudin A A, M\'endez E R and Shchegrov A V 2000 
Multiple-scattering phenomena in the scattering of light from 
randomly rough surfaces in
{\it Frontiers of Laser Physics and Quantum Optics}, ed Z Xu, S Xie, S
Y Zhu and M O Scully (New York : Springer-Verlag) pp 225-259
\item{[8]}
Maradudin A A, M\'endez E R and Shchegrov A V (unpublished work)
\item{[9]}
Press W H, Teukolsky S A, Vetterling W T and Flannery B P 1992
{\it Numerical Recipes in Fortran, 2nd ed} (New York : Cambridge 
University Press) pp 281-282
\item{[10]}
Maradudin A A, Michel T, McGurn A R and M\'endez E R 1990
Enhanced backscattering of light from a random grating
{\it Ann. Phys. (N.Y.)} {\bf 203} 255-307
\end{description}
\end{document}